\documentclass{hotnets17}
\usepackage[normalem]{ulem}
\usepackage{times}  
\usepackage{hyperref}
\usepackage{color}
\usepackage{enumitem}
\usepackage{dsfont}
\usepackage{xspace}
\usepackage{graphicx}
\usepackage[margin=5pt,font=small,labelfont=bf]{subfig}

\usepackage{floatrow}

\usepackage{microtype}
\usepackage[font=small,labelfont=bf]{caption}
\hypersetup{pdfstartview=FitH,pdfpagelayout=SinglePage}

\newcommand{\noteav}[1]{{\color{magenta} (AV: #1)}}

\newcommand{\miniheader}[1]{\miniheaderclean{#1.}}
\newcommand{\miniheaderclean}[1]{{ \vspace{0.05in}\noindent {\bf #1} }}

\newcommand{\rs}{$\mathcal{R}$\xspace}
\newcommand{\rsimple}{$\mathcal{RS}$\xspace}
\newcommand{\rst}[1]{$\mathcal{R}^{#1}$}

\newcommand\eat[1]{}

\begin{document}




\title{A Machine Learning Approach to Routing}


\author{Asaf Valadarsky$^1$
  \and Michael Schapira$^1$
    \and Dafna Shahaf$^1$
  \and Aviv Tamar$^2$ \\
  \and $^1$School of Computer Science and Engineering, The Hebrew University of Jerusalem, Israel\\
    $^2$Dept. of Electrical Engineering and Computer Sciences, UC Berkeley, USA\\[2ex]
}

\maketitle


\begin{abstract}
Recently, much attention has been devoted to the question of whether/when traditional network protocol design, which relies on the application of algorithmic insights by human experts, can be replaced by a data-driven (i.e., machine learning) approach. We explore this question in the context of the arguably most fundamental networking task: routing. Can ideas and techniques from machine learning (ML) be leveraged to automatically generate ``good'' routing configurations? We focus on the classical setting of intradomain traffic engineering. We observe that this context poses significant challenges for data-driven protocol design. Our preliminary results regarding the power of data-driven routing suggest that applying ML (specifically, deep reinforcement learning) to this context yields high performance and is a promising direction for further research. We outline a research agenda for ML-guided routing.


\end{abstract}

\maketitle

\section{Introduction}

Applying machine learning (ML) to computational challenges is prevalent in numerous areas in computer science (AI, computer vision, graphics, NLP, comp-bio, and beyond). Computer networking, in contrast, has largely withstood the ML tide until recently. Recent advances suggest that this might be changing~\cite{data-driven,pensive,deeprm}.

We ask whether data-driven protocol design~\cite{data-driven} can improve upon today's approaches in the context of routing, i.e., the selection of paths for traffic within a network, or across networks. 

\miniheaderclean{Why apply machine learning to routing?} Routing is, arguably, the most fundamental networking task and, consequently, has been extensively researched in a broad variety of contexts (data centers, WANs, ISP networks, interdomain routing with BGP, wireless networks, and more). Traditionally, route-optimization contends with uncertainty about future traffic conditions in one of two manners: (1) optimizing routing configurations with respect to \emph{previously observed} traffic conditions, with the hope that these configurations fare well also with respect to the future, or (2) optimizing with respect to \emph{a range} of feasible traffic scenarios, in hope of providing high performance across the entire range~\cite{Fortz,coyote,kulfi,oblivious}. 

Unfortunately, in general, routing configurations optimized with respect to specific traffic conditions can fail miserably in achieving good performance even under not-too-different traffic conditions. In addition, optimizing worst-case performance across a broad range of considered traffic scenarios might come at the expense of being far from the achievable optimum for the actual traffic conditions.

Intuitively, ML suggests a third option: leveraging information about \emph{past} traffic conditions to learn good routing configurations for \emph{future} conditions. While the exact future traffic demands are unknown to the decision maker in advance, a realistic assumption is that the history of traffic demands contains \emph{some} information regarding the future (e.g., changes in traffic across times of day, the skewness of traffic, whether certain end-hosts communicate often, etc.). Hence, a natural approach is to continuously observe traffic demands and adapt routing with respect to (implicit or explicit) \emph{predictions} about the future. 



\miniheaderclean{Intradomain traffic engineering (TE) as a case study.} We initiate the study of ML-guided routing by examining the classical environment of intradomain TE~\cite{RexfordSurvay,Fortz,daytime,tightrope,peft,kulfi,coyote}--the optimization of routing within a single, self-administered network. We leave the investigation of data-driven routing in other contexts to future research (Section~\ref{sec:conclusion}).

We present a model for data-driven (intradomain) routing that builds on the rich body of literature on intradomain TE~\cite{RexfordSurvay,Fortz,daytime,tightrope,peft,kulfi,coyote} and (multicommodity~\cite{multicommodity-throughput,multicommodity,oblivious,coyote,kulfi,Fortz,multicommodity-fairness}) flow optimization. We investigate, within this model, the application of different ML paradigms and machinery. 

In our investigation of ML-guided intradomain TE we grapple with two main questions: (1) How should routing be formulated as an ML problem? and (2) What are suitable representations for the inputs and outputs of learning in this domain? We next expand on each of these challenges, which also pertain to data-driven routing in other contexts.




\miniheaderclean{Learn future traffic demands or learn routing configurations? Supervised learning or reinforcement learning?} A natural approach to ML-based routing is the following: observe past traffic demands, apply ML to \emph{explicitly} predict the upcoming traffic demands, and optimize routing with respect to the predicted demands. In ML terms, this is a \emph{supervised learning} task~\cite{book-sl}.

We evaluate several supervised learning schemes for predicting traffic demands. Our preliminary results are discouraging, indicating that supervised learning might be ineffective if the traffic conditions do not exhibit very high regularity. We next turn our attention to a different approach: \emph{reinforcement learning}~\cite{rl-book}. Now, instead of explicitly learning future traffic demands and optimizing with respect to these, the goal is to learn a good mapping from the observed history of traffic demands to routing configurations. Our preliminary results suggest that this approach is more promising, yet realizing it requires care, as discussed next.

\miniheaderclean{What should the output of the learning scheme be?} The intradomain routing context poses significant challenges to the application of reinforcement learning. A key challenge is that the natural ``output'' of a routing scheme is a collection of rules specifying how traffic is forwarded from each source to each destination. This output's naive representation involves a very large set of parameters (as opposed to, e.g., selecting a single action from a fairly small set~\cite{deeprm,pensive}). Our initial results indicate that this can render learning slow and ineffective. We hence devise methods for constraining the size of the output without losing ``too much'' in terms of routing expressiveness. We leverage ideas from the literature on hop-by-hop traffic engineering~\cite{halo,peft,Fortz} to efficiently learn, via deep reinforcement learning, good routing configurations. Our preliminary findings suggest that this is a promising direction for improving upon today's intradomain TE.

\miniheaderclean{Outlining a research agenda for data-driven routing.} We believe that our investigation below but scratched the surface of data-driven routing. We leave the reader with many interesting research questions, including (1) extending our approach to other routing contexts, (2) examining other performance metrics, (3) identifying better supervised learning approaches to traffic-demand estimation, (4) scaling reinforcement learning in this context, and beyond. We discuss this research agenda in Section~\ref{sec:conclusion}.


\section{Data-Driven Routing Model}\label{sec:model}

In our framework, a decision maker (network operator / automated system) repeatedly selects routing configurations. Traffic conditions vary and routing decisions are oblivious to future traffic demands. Our focus is on the conventional optimization objective of minimizing link over-utilization (a.k.a. minimizing congestion) from traffic engineering literature~\cite{Fortz,oblivious,coyote,kulfi}.

\miniheader{Network} We model the network as a capacitated directed graph $G=(V,E,c)$, where $V$ and $E$ are the vertex and edge sets, respectively, and $c:E \to \mathbb{R^+}$ assigns a capacity to each edge. Let $n$ denote the number of vertices in $V$ and $\Gamma(v)$ denote vertex $v$'s neighboring vertices in $G$. 

\miniheader{Routing} A \emph{routing strategy} \rs for the network specifies, for each source vertex $s$ and destination vertex $t$ how traffic from $s$ to $t$ that traverses $v$ is split between $v$'s neighbors. Thus, a routing strategy specifies, for each vertex $v$ and source-destination pair $(s,t)$ a mapping from $v$'s neighbors to values in the interval $[0,1]$, $\mathcal{R}_{v,(s,t)}:\Gamma(v) \to [0,1]$, such that $\mathcal{R}_{v,(s,t)}(u)$ is the fraction of traffic from $s$ to $t$ traversing $v$ that $v$ forwards to its neighbor $u$. We require that for every $s,t\in V$ and $v\neq t$, $\sum_{u \in \Gamma(v)} \mathcal{R}_{v,(s,t)}(u) = 1$ (no traffic is blackholed at a non-destination), and also for every $s,t\in V$, $\sum_{u \in \Gamma(v)} \mathcal{R}_{t,(s,t)}(u) = 0$ (all traffic to a destination is absorbed at that destination).

\miniheader{Induced flows of traffic} A \emph{demand matrix} (DM) $D$ is a $n\times n$ matrix whose $(i,j)$'th entry $D_{i,j}$ specifies the traffic demand between source $i$ and destination $j$. Observe that any demand matrix $D$ and routing strategy \rs induce a flow of traffic in the network, as explained next. Traffic from every source $s$ to destination $t$ is split amongst $s$'s neighbors according to $\mathcal{R}_{s,(s,t)}$. Similarly, traffic from $s$ to $t$ traversing a neighbor of $s$, $v$, is split amongst $v$'s neighbors according to $\mathcal{R}_{v,(s,t)}$, etc.

\miniheaderclean{How good is a traffic flow?} We adopt the classical objective function of minimizing link (over)utilization~\cite{Fortz,coyote,oblivious,kulfi}. The link utilization under a specific multicommodity flow $f$ is \emph{max}$_{e \in E} \frac{f_e}{c(e)}$, where $f_e$ is the total amount of flow traversing edge $e$ under flow $f$.  Our formulation can easily be extended to other multicommodity-flow-based objective functions. We leave the evaluation of other objectives (e.g., flow-completion time, latency) to future research (Section~\ref{sec:conclusion}).

We point out that for any \emph{given} demand matrix, computing a multicommodity flow $f$ that minimizes link utilization can be executed in a computationally-efficient manner via linear programming~\cite{multicommodity,oblivious,Fortz}. Our focus, in contrast, is on the realistic scenario in which the DM is not known beforehand.

\miniheader{Routing future traffic demands} Time is divided into consecutive intervals, called ``epochs'', of length $\delta_t$ ($\delta_t$ is determined by the network operator). At the beginning of each epoch $t$, the routing strategy \rst{t} for that epoch is decided. \rst{t} can depend only on the history of \emph{past} traffic patterns and routing strategies (from epochs $1,\ldots,t-1$). 

We make two simplifying assumptions: (1) the demand matrix is fixed throughout each time epoch, and (2) demand matrices can be inferred after the fact (e.g., via network measurements). We leave the investigation of data-driven routing under more complex traffic patterns (e.g., IP flows enter and leave within each epoch) and of information-constrained routing decisions (e.g., only partial information about past traffic demands) to the future.

After selecting the routing strategy \rst{t} for epoch $t$, the demand matrix for epoch $t$, and the associated cost, in terms of maximum link-utilization, are revealed. The objective of the decision maker is to select routing strategies in a manner that consistently results in low link over-utilization.

\section{What to Learn?}

\begin{figure*}[t]
    \centering
    
    \subfloat[\em Cyclic gravity DM sequences (sparsity $p=0.3$)\label{fig:sl-success-gravity}]{
      \includegraphics[width=0.30\linewidth, height=0.18\textheight, keepaspectratio=true]{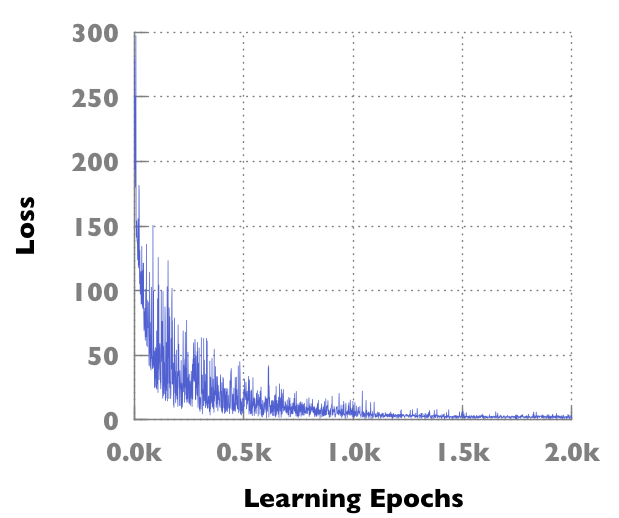}
    }\hspace{2em}%
    \subfloat[\em Averaged bimodal DM sequences (sparsity $p=1$)\label{fig:sl-success-bimodal}]{
        \includegraphics[width=0.30\linewidth, height=0.18\textheight, keepaspectratio=true]{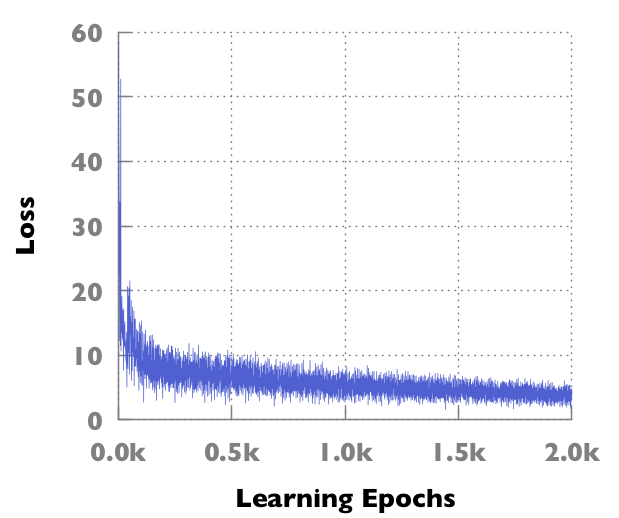}
    }\hspace{2em}%
    \subfloat[\em Randomly drawn gravity DM sequences (sparsity $p=0.3$)\label{fig:sl-failure}]{
  		\includegraphics[width=0.30\linewidth, height=0.18\textheight, keepaspectratio=true]{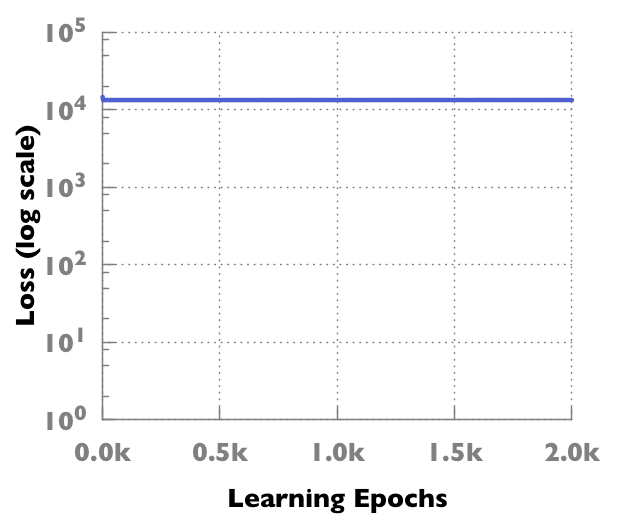}
  	}
    
    \caption{\em Representative Results for Supervised Learning (using \emph{NAR-NN} with $k=10$ and $q=5$)}\label{fig:sl}
\end{figure*}

Our underlying assumption is the existence of some \emph{regularity} in the DMs, and the purpose of the investigation below is exploring how such regularity can be inferred and leveraged to optimize routing. We consider two different manifestations of regularity--- embedding \emph{deterministic} regularity into the DM sequence and drawing DMs from a \emph{fixed probability} distribution---and two high-level learning approaches---supervised learning and reinforcement learning.

\subsection{Supervised Learning Approach}\label{sec:supervised}

Since for a given demand matrix (DM), an optimal routing strategy is efficiently computable, a natural approach is to repeatedly try to predict (i.e., learn) the next DM and then compute an optimal routing strategy for that DM. In ML terms, this is a supervised learning problem.

\miniheader{Supervised learning} A supervised learning task involves a sample space $\mathcal{X}$ and a labeling space $\mathcal{Y}$. An algorithm $\mathcal{A}$ for the task is a function mapping values in $\mathcal{X}$ to labels in $\mathcal{Y}$. Given a set of samples and their true labels $\{(x_i,y_i)\} \in \mathcal{X} \times \mathcal{Y}$, the goal is to identify a mapping that 
produces correct labels for new samples, drawn from the same distribution as the data. How good/bad a mapping fares is quantified in terms of a loss function $\mathcal{L} : \mathcal{Y} \times \mathcal{Y} \to \mathbb{R}$. Intuitively, for any pair of labels $(y_1,y_2)$, $\mathcal{L}(y_1,y_2)$ represents the cost of predicting, for a given sample, the label $y_2$ instead of the correct label $y_1$. See~\cite{book-sl} for a detailed exposition of supervised learning.

We consider the following supervised learning approach to routing: the learning algorithm observes the history of DMs up to the current epoch, and predicts the DM for the upcoming epoch. This prediction is then used to generate an optimal routing strategy with respect to the predicted DM. When is employing this scheme a good idea? To answer this question, we evaluate several supervised learning schemes for predicting the next DM on different traffic patterns.

\miniheader{Generating DM sequences for our experiments} We next discuss how traffic patterns are generated in our experiments. We consider two standard schemes for generating DMs: the (deterministic) gravity model~\cite{gravity} and the (probabilistic) bimodal model~\cite{bimodal}. Intuitively, the former captures scenarios in which communication between end-points is proportional to their outgoing bandwidths and the latter captures scenarios in which communication end-points are divided into small flows (mice) and large flows (elephants). 
\eat{
See~\cite{gravity,bimodal} for details. 
}
We also consider ``\emph{sparsifications}'' of gravity/bimodal DMs generated by selecting, uniformly at random, a $p$-fraction of the communicating pairs, for some choice of $p\in [0,1]$, and removing the traffic demands of all other pairs from consideration. We refer to $p$ as the \emph{sparsity} of the DM.

Our experiments require generating \emph{sequences} of DMs, specifying a DM for each time epoch. We examine two classes of DM sequences:


\miniheader{Class I: DM sequences in which the next DM is \emph{deterministically} derived from past DMs} One example for such a DM sequence is ``a cycle of DMs'', in which the DM in each epoch belongs to a fixed set of $q$ DMs, $D^{(0)},\ldots,D^{(q-1)}$ such that if $D^{(j)}$ is the DM in epoch $t-1$ then $D^{(j+1\ modulo\ q)}$ is the DM in epoch $t$. $D^{(0)},\ldots,D^{(q-1)}$ in our experiments are sparsified (for varied values of $p$) gravity/bimodal DMs (for varied values for parameters of the bimodal model). Cycles of DMs might capture, e.g., the scenario that the traffic demands at a certain time of day are rather similar across days. See discussion of such temporal consistencies in ISP networks in~\cite{daytime}. Another example of a DM sequence that, though more artificial, also exhibits high regularity (and so is interesting to study) is when each DM is the average of the previous $q$ DMs (for some fixed $q>0$). Our experiments evaluate supervised learning schemes on DM cycles of sizes $q=5,10,15,20$, and DM sequences in which each DM is the average over the previous $q=5,10,15,20$ DMs.

\miniheader{Class II: DM sequences in which each DM is independent of the previous DMs} The DM for each epoch is now drawn independently from a fixed probability distribution over DMs, namely sparsified gravity/bimodal DMs. We point out that such traffic patterns are commonly used in evaluations of data center architectures and protocols~\cite{fatfree,dctcp,augmenting,mirror-mirror} as traffic in data centers is often viewed as highly skewed and unpredictable~\cite{firefly,projector}.


\miniheader{Supervised learning schemes} Following the recent successes of deep neural networks (DNNs)~\cite{krizhevsky2012imagenet,mnih2015human,schulman2015trust}), we evaluate 3 different DNN architectures. The input to all three architectures is the $k$ most recent observed DMs and the output is a DM. We examine different values of $k$ ($5$, $10$, and $20$). We use the Frobenius (or $l_2$) norm~\cite{l2_loss} to quantify the quality of an output with respect to the actual DM.  
The three architectures differ in the structure of the neural network interconnecting the input layer (representing $k$-long histories of DMs) and output layer (representing the next DM). We evaluate (1) \emph{FCN}, a 3-layered fully-connected network, (2) \emph{CNN}, a 4-layered convolutional neural-network~\cite{cnn}, and (3) \emph{NAR-NN}, a nonlinear auto-regressive model~\cite{nar}, realized via a 4-layered neural network that, for input demand matrices $D^{(1)},\ldots,D^{(k)}$, learns a $k$-vector $\alpha=(\alpha_1,\ldots,\alpha_k)$ and an $n\times n$ matrix $\beta$, and outputs the DM $\sum_i \alpha_i D^{(i)} + \beta$.

\miniheader{Evaluation framework} We experiment with gravity and bimodal DMs of various sizes ($9\times 9$, $12 \times 12$, $23 \times 23$, $30\times 30$, $50\times 50$, and $100\times 100$) under various choices of sparsity levels $p=0.3,0.6,0.9,1$ and of values of per vertex outgoing bandwidths (ranging from 10's of MB to 10's of GB). We consider various DM sequence lengths for training and testing the model (ranging from a few 10's to few 100's of DMs). We generate, for each choice of parameter assignment to $q,p,k$ and sequence length, a training set of 10 DM sequences and a test set of 3 DM sequences. We define a \emph{learning epoch} as a full traversal of the training set. We train each neural network for $2000$ learning epochs.

    
  	

\miniheader{Results} Our experimental results (for the test DM sequences) show that for DM sequences that exhibit deterministic regularity, namely, cycles of DMs and ``averaged DMs'', only the \emph{NAR-NN} performs fairly well and only for specific relations between the examined history ($k$) and the size of the cycle / number of DMs averaged over ($q$). Specifically, when $q \leq k$, \emph{NAR-NN} well-approximates the next DM for cycles of DMs, and performs well on averaged DMs. \emph{NAR-NN} continues to perform reasonably well on averaged DMs when $q>k$, but fails on cycles of DMs for $q>k$. All 3 architectures failed to approximate the next DM for randomly generated DMs (which is not surprising, as there are no temporal correlations between DMs in the sequence). 

We present representative results for \emph{NAR-NN} on a network $G$ with $30$ vertices. We plot the loss, in terms of distance of the predicted DM from the actual DM (y-axis), over the number of learning epochs (x-axis). 
Figure~\ref{fig:sl-success-bimodal} and Figure~\ref{fig:sl-success-gravity} show that the model succeeds in learning the next DM when using the averaged and cyclic DM sequence generation. Figure~\ref{fig:sl-failure} demonstrates failure in learning the next DM when drawn from a probability distribution. We leave the investigation of whether better supervised learning of traffic demands is feasible for the future (see Section~\ref{sec:conclusion}).

\subsection{Reinforcement Learning Approach}

Next, instead of learning the next DM and optimizing the routing strategy with respect to that DM, our goal is to directly learn a good mapping from observed DMs to routing strategies.

\miniheader{Reinforcement learning} In the reinforcement learning framework, an agent repeatedly interacts with an environment. Time is divided into discrete time slots $t=1,2,3,...$. At the beginning of each time slot $t$, the agent observes the current \emph{state} $s_{t-1}$ of the environment and selects an action $a_t$ from a fixed set of actions. Once the agent chooses action $a_t$, the state of the environment changes to $s_t$ and the agent receives a \emph{reward} $r_t$ (a numerical value) signifying how good/bad the action he took was. The goal of the agent is to learn a mapping $\pi$ from the set of possible states $\mathcal{S}$ to the space of actions $\mathcal{A}$ (i.e., $\pi: \mathcal{S} \to \mathcal{A}$) that fares well with respect to the objective of maximizing the \emph{expected discounted reward} $\mathbb{E}[\sum_{t} \gamma^t r_t]$ for a predetermined $\gamma>0$, called the \emph{discount factor}. See~\cite{rl-book} for a detailed exposition of reinforcement learning.

\miniheader{Routing via reinforcement learning} Routing-strategy selection can be cast as a reinforcement learning task as follows. At the beginning of each time epoch $t$, the operator/system (agent) decides on a routing strategy \rst{t} for that epoch based on the routing strategies and DMs in the most recent $k$ time epochs, which constitute the observed state of the environment at that point. Then, the state changes as the DM for epoch $t$, $D^{(t)}$, is revealed and the reward $r^{(t)}=-\frac{u^{(t)}}{OPT^{(t)}}$ is received, where $u^{(t)}$ is the max-link-utilization under \rst{t} for $D^{(t)}$, and $OPT^{(t)}$ is the optimal max-link-utilization with respect to $D^{(t)}$ ($r^{(t)}$ thus captures the ratio between achieved performance and optimal performance). The goal is to learn a mapping from $k$-long histories of DMs to routing strategies that maximizes the expected discounted reward, as formulated above. We explore the power of this approach in the following sections.

\section{Representing the Output}\label{sec:splitting}

In contrast to other recent applications of ML to networking~\cite{deeprm,pensive}, learning routing strategies involves generating neural networks with very large output layers (containing, e.g., thousands of output nodes even for a communication network of but tens of vertices). Consider, e.g., the representation of a routing strategy described in Section~\ref{sec:model}. This representation involves $|V|^2\cdot|E|$ variables, where $|V|$ and $|E|$ are the sizes of the network graph's vertex set and edge set, respectively. We show below that even for constrained routing strategies of much smaller sizes, a (``vanilla'') reinforcement learning approach for predicting the complete routing strategy fails to attain good performance within reasonable time.

We restrict our attention to \emph{destination-based} routing strategies, i.e., routing strategies in which the splitting ratios at each vertex $u$ with respect to any destination $d$ are the same across all possible sources $s$. Observe that any such routing strategy \rsimple can be represented by $|V| \cdot |E|$ values (i.e., $|V|$ times smaller than unconstrained routing strategies). 
We employ the continuous-control reinforcement-learning algorithm, TRPO~\cite{schulman2015trust}, applied to a 3-layered fully-connected neural network, to learn the mapping $\pi$ from $k$-long histories of DMs to a routing strategy \rsimple. The real-valued outputs generated by the deep neural network are turned into per vertex traffic-splitting ratios by applying, for each vertex in the communication network $u$, the softmax function~\cite{softmax} to the outputs corresponding to $u$'s outgoing edges.

\miniheader{Evaluation} We adapt the open-source implementation of TRPO~\cite{schulman2015trust} provided by OpenAI~\cite{trpo-github} to the task of learning routing-strategies. We begin our evaluation with a seemingly easy target: learning the splitting ratios for a 12-vertices, 32-edges network (taken from~\cite{topozoo}), and (sparsified) gravity DMs. We train a 3-layered fully-connected network over $7$ sequences of gravity DMs of length $60$ and evaluate (test) the neural network on $3$ such sequences. We repeat this process for sparsity levels $0.3,0.6$, and $0.9$. We use $k=10$ (the length of the history of past DMs received as input). We compute the optimal congestion using the CPLEX~\cite{cplex} LP solver. 

The training phase involves generating, from every sequence of DMs of length $60$, $50$ sequences of $10$ consecutive DMs (representing ten-long histories of DMs), by grouping DMs $1-10$, $2-11$, etc. Training the neural network on each of these ``histories of DMs'' involves evaluating the neural network $30$ times in parallel  (and so $1,500$ iterations per DM sequence and $10,500$ overall). We refer to one execution of this process as a ``learning epoch''. 

\eat{
As seen in Table~\ref{tbl:splitting_vs_w_learning}, even after many learning epochs, the resulting routing strategies are still very far with respect to the optimum, e.g., after \noteav{$???$} learning epochs (which take approximately \noteav{???} days to compute) for sparsity value $0.9$, the produced routing strategies are still \noteav{$???$x} away from the optimum, in terms of max-link-utilization. As shown in Section~\ref{sec:softmin_routing}, routing strategies that fare significantly better can be generated much quicker.
We conclude that the number of output parameters renders efficient learning very challenging (at best). We next discuss how the output size can be significantly decreased without compromising on good routing outcomes.
}

Our results (omitted due to space constraints) suggest that this approach leads to slow and ineffective learning; e.g., even after more than $700$ learning epochs, the produced routing strategies were still over 9x away from the optimum, in terms of max-link-utilization. As shown below, routing strategies that fare significantly better can be generated much quicker. We hypothesize that the large number of output parameters renders efficient learning very challenging. We thus seek a class of routing policies that can be more concisely represented yet is still rich enough to attain high performance.

    


\section{Learning Softmin Routing}\label{sec:softmin_routing}

\begin{figure}[t]
    \centering
    \subfloat[\em Congestion ratio for sparse ($p=0.3$) gravity DM sequences\label{fig:rl-gravity-sparse-ob}]{
      \includegraphics[width=0.65\linewidth, height=0.33\textheight, keepaspectratio=true]{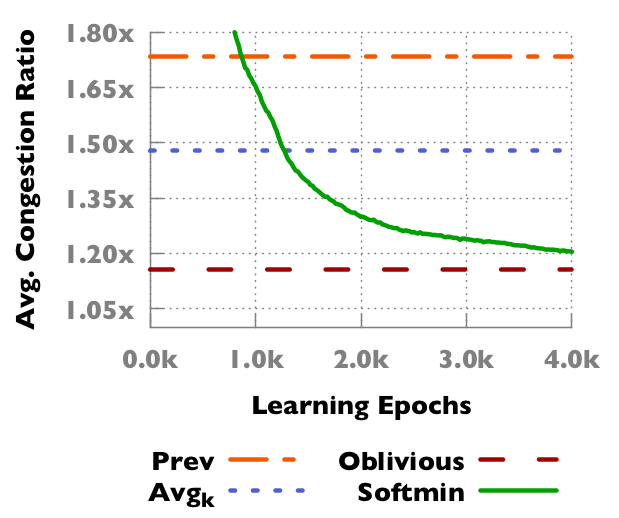}
    }
    
    \subfloat[\em Congestion ratio for non-sparse ($p=1.0$) bimodal DM sequences with 40\% elephant flows\label{fig:rl-bimodal-medium-ob}]{
  		\includegraphics[width=0.65\linewidth, height=0.33\textheight, keepaspectratio=true]{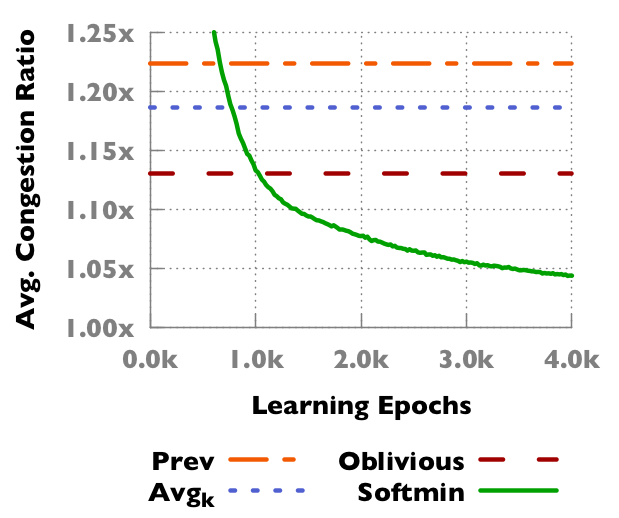}
  	}
    \caption{\em Representative Results for $softmin$-Routing}\label{fig:rl-ob}
\end{figure}

We explore the following approach: instead of learning splitting ratios directly, learn \emph{per-edge weights}, and then use these weights to generate a routing strategy. Under this approach, the output of the neural network is of size $|E|$ (as opposed to $|V|^2\times |E|$ and $|V|\times |E|$ for unrestricted and destination-based routing policies, respectively).

Generating forwarding rules from link weights is a classical approach to routing~\cite{Fortz,peft,halo}. We resort to the following approach:

The $softmin_\gamma$ value for a vector of $r$ coordinates $\alpha=(\alpha_1,\alpha_2,\dots,\alpha_r)$, for $\gamma>0$, is the vector (also of $r$ coordinates)
$softmin_\gamma(\alpha)_i = \frac{e^{-\gamma \alpha_i}}{\sum_{i=1}^r e^{-\gamma \alpha_i}},\quad i\in 1,\dots,r.$
Observe that $softmin(\alpha)$ can be regarded as a probability distribution (as the sum of all coordinates necessarily equals $1$). 


Consider a specific assignment of per-edge weights $w=\{w_e\}_{e\in E}$, a specific edge $(u,v)\in E$, and a specific traffic destination $d$. Observe that, when viewing weights as distances, $w$ determines the length of the shortest path from vertex $u$ to vertex $d$ that goes through $u$'s immediate neighbor $v$. Let $SP_w(v,u,d)$ denote this length. Given a set of such per-neighbor distances for a vertex $u$, the $softmin$ function can be applied to generate a probability distribution across these neighbors, which can be interpreted as $u$'s splitting ratios for traffic destined for $d$. We refer to this scheme as ``\emph{softmin-routing}''. The higher the choice of $\gamma$ to plug into the softmin function the closer the resulting routing scheme is to shortest path routing. We set $\gamma=2$ in our experiments. 

Our reinforcement learning scheme maps $k$-long histories of DMs to per-edge link weights. The reward is computed by turning these weights into traffic splitting ratios and computing the max-link-utilization of the resulting routing strategy with respect to the next DM. We realize this learning scheme via a 3-layers fully-connected network.

We benchmark our results against three alternative non-ML-based approaches to computing routing strategies: (1) $Prev$: optimizing softmin routing with respect to the most recent DM, (2) $Avg_k$, optimizing softmin routing with respect to the $k$ most recent DMs, and (3) $Oblivious$, the optimal oblivious routing scheme~\cite{oblivious} (which does not depend on the history of DMs at all).
\footnote{Observe that both $Prev$ and $Avg_k$ optimize softmin routing, as opposed to applying the optimal multicommodity flow computed for the input DM(s) to route the next DM. The reason is that the latter option is not well defined (and, in particular, some of the end-points communicating in the next DM might not communicate at all in the input DM). We point out that in our experimentation softmin routing is consistently within at most $5\%$ of the optimum traffic flow with respect to any input DM and so very closely approximates this strategy.}


    

\miniheader{Evaluation} We consider a communication network with $12$ vertices and $32$ edges.
We use the adaptation of~\cite{trpo-github} discussed in Section~\ref{sec:splitting} to train a 3-layered fully-connected neural network to generate the weights for softmin routing. We train the neural network on $7$ sequences of gravity and bimodal DMs of length $60$ each, and tested on $3$ such sequences. For gravity DM, the above process is repeated per sparsity levels $0.3,0.6$, and $0.9$. For bimodal DMs the sparsity level is $p=1$ and the percentage of large (elephant) flows amongst the communicating pairs is varied: $20/40/60\%$ of all pairs. We set $k=10$ and compute the optimal flow via the CPLEX~\cite{cplex} LP solver. 

We show in Figure~\ref{fig:rl-gravity-sparse-ob} representative results for gravity DMs and in Figure~\ref{fig:rl-bimodal-medium-ob} representative results for bimodal based DMs. The figures plot the ratio between the performance of the resulting routing strategies (for the test DM sequences), in terms of averaged max-link-utilization (congestion), and the optimum congestion.
Interestingly, oblivious routing outperforms the other two baselines. Observe that $softmin$-routing gets very close to oblivious routing's performance for gravity DM sequences (and could perhaps outperform it with more training), and significantly outperforms oblivious routing for bimodal DM sequences.

\section{Related Work} 

    



\noindent{\bf Traffic engineering}  Traffic engineering (TE) is fundamental to networking, and hence vastly researched. Results on TE range from routing in legacy, OSPF/ECMP networks~\cite{Fortz,coyote} to datacenter networks~\cite{hedera} and backbone networks~\cite{tightrope}. Softmin routing is inspired by the literature on TE via hop-by-hop routing in IP networks e.g., PEFT~\cite{peft} and HALO~\cite{halo}. We find softmin routing especially convenient to use as it involves fairly simple splitting traffic across next-hops, while still achieving high performance.

\miniheader{Reinforcement learning} Machine learning via deep-neural networks has proven extremely useful in executing many different tasks: machine translation~\cite{bahdanau2014neural}, image recognition~\cite{krizhevsky2012imagenet}, and more. Specifically, reinforcement learning has been applied to playing computer games~\cite{deepmind-atari} and beating world-champions in strategic board games~\cite{alphago}, robotics~\cite{rl-robotics}, 3D-locomotion tasks~\cite{schulman2015trust}, and beyond. The development and optimization of reinforcement learning algorithms is thus the subject of much attention. Our algorithms rely on utilizing TRPO~\cite{schulman2015trust}. We leave the evaluation of other reinforcement-learning algorithms~\cite{a3c,acktr,ppo} to future research.

\miniheader{ML applications to networking} Machine learning has been applied to various networking contexts including congestion control~\cite{remy,pcc}, network bottleneck detection~\cite{lube}, and optimizing datacetner power consumption~\cite{google-cooling-dc}, resource allocation~\cite{deeprm}, and bitrate selection for video streaming~\cite{pensive}. Q-routing~\cite{q-routing} applies Q-learning~\cite{q-learning} to the network routing context. Under Q-routing~\cite{q-routing}, each router individually learns a mapping from packet headers to outgoing ports. This involves routers constantly exchanging information, at per packet resolution, about their latencies with respect to different destinations. We believe that operating at per packet level, and in a decentralized fashion, poses significant challenges in terms of scalability and communication overhead.

\section{Conclusion}\label{sec:conclusion}

We initiated the study of data-driven routing and presented preliminary results for the context of intradomain traffic engineering. Our preliminary results from experimentation with deep reinforcement learning show that extracting information from the history of traffic scenarios to generate good routing with respect to future traffic scenarios is an interesting approach. We view our results as a first step towards realizing a much broad research agenda. 

\miniheaderclean{Other routing domains.} Optimizing routing is a keystone of networking research, investigated in a broad variety of contexts, including legacy IP networks~\cite{Fortz}, data centers~\cite{hedera}, private backbone networks~\cite{tightrope}, interdomain routing with BGP, overlay networks~\cite{ron}, publish-subscribe networks~\cite{publish-subscribe}, and more. Applying a data-driven routing approach to other settings is an important research direction.

\miniheaderclean{Other objective functions.} Our focus in this study was on the classical objective of minimizing max-link-utiliziation. Examining other well-studied multicommodity-flow-based objectives, e.g., maximizing overall goodput, is of great interest, as is investigating performance metrics that relate to latency, flow-completion-time, etc.

\miniheaderclean{Predicting traffic-demands.} Our preliminary results suggest that well-predicting traffic conditions can, in general, be very challenging. This motivates further research on supervised learning approaches to this challenge.

\miniheaderclean{Better ML-guided intradomain traffic engineering.} Our investigation of the application of ML to intradomain TE is only a first step in this direction. Important questions remain regarding (1) the scalability of ML approaches in this context, (2) the environments in which ML-guided routing outperforms traditional routing, and the causes for this, and (3) the ``right'' choice of the duration of the time epoch be to strike the right balance between routing stability and reactiveness to traffic changes.

\miniheaderclean{Better experimental and empirical evaluations.} Our experiments involved fairly small networks and synthetically-generated traffic demands. Evaluating routing solutions in more realistic scenarios is important.

\section*{Acknowledgements}
We thank Marco Chiesa for providing us with the code needed to evaluate the oblivious routing scheme. We also thank the anonymous HotNets reviewers for valuable feedback.

\end{document}